\documentclass[twocolumn,showpacs,aps,prd,amsmath,amssymb,floatfix,nofootinbib]{revtex4}

 \usepackage{graphicx}
 \usepackage{dcolumn}
 \usepackage{bm}
 \usepackage{indentfirst}
 \usepackage{amssymb}
 \usepackage{epsfig}
 \usepackage{latexsym}
 \usepackage{indentfirst}
 \usepackage[english]{babel}
 \usepackage[latin1]{inputenc}
 \usepackage{amsmath}
\usepackage{hyperref}



\newcommand{\beq}{\begin{equation}}
\newcommand{\eeq}{\end{equation}}
\newcommand{\beqa}{\begin{eqnarray}}
\newcommand{\eeqa}{\end{eqnarray}}

\newcommand{\be}{\begin{equation}}
\newcommand{\ee}{\end{equation}}
\newcommand{\ben}{\begin{equation*}}
\newcommand{\een}{\end{equation*}}
\newcommand{\bea}{\begin{eqnarray}}
\newcommand{\eea}{\end{eqnarray}}
\newcommand{\besin}{\begin{eqnarray*}}
\newcommand{\eesin}{\end{eqnarray*}}

\begin{document}

\title{Non Linear Lorentz Transformation  and Doubly Special Relativity}

\author{A. N. Atehortua $^{1,2}$, D. E. Jaramillo $^1$,
  J. M. Mira $^1$, N. Vanegas $^1$ }

\thanks{Email: danielj@fisica.udea.edu.co}

\affiliation{
$^1$Instituto de F\'{i}sica, Universidad de Antioquia, AA 1226, Medell\'{i}n, Colombia.\\
$^2$Facultad de Ciencias, Instituto Metropolitano de Medell\'{i}n, \\Calle 73 No 76A-354 Medell\'{i}n, Colombia.}
\date{\today}


\begin{abstract}
  We   generate non-linear representations of the Lorentz
  Group by unitary transformation over the Lorentz generators. To do that we use  
 deformed scale transformations by introducing  momentum-depending parameters. 
 The momentum operator transformation is found to be equivalent to a particle momentum 
transformation. The configuration space transformation is found to  
depend on the old momentum operator and we show that this transformation generates models
with two scales, one for the velocity ($c$) and another one for the energy. 
A Lagrangian formalism is proposed for these models 
and an effective metric for the  deformed Minkowski space is found. We show 
that the Smolin model is one in a family of doubly special relativity.
Finally we construct an ansatz for the quantization of such theories.  

\end{abstract}

\maketitle

\section{Introduction}

There are several theoretical reasons to establish a new physical and
fundamental scale of nature that would enable us to explore deep into
the transition threshold between general relativity and quantum
mechanics \cite{Smolin:2006pa}, \cite{PhysRevLett.88.190403}. 
One of those reasons is to come closer to a
theory of quantum gravity $QG$ \cite{rovelli}. Traditionally, the
Planck's length $L_{p}=\sqrt{\hbar G/c^3}$ has been postulated as a
fundamental scale. However, there are other ways to explore this
second invariant scale. These are called deformed special relativity
(DSR) models and can considered a theoretical limit of some larger
theories  of Quantum Gravity (QG)
\cite{Smolin:2010xa}. DSR models are being outlined as a
phenomenological description of presently unknown quantum gravity
effects \cite{Calmet:2010tx}. Those models are initially thought as a
way of modifying the dispersion relations of relativistic particles
without introducing a preferred frame. This modifications have been
used to look into CPT/Lorentz symmetry violations \cite{Dolgov:2009yk} and
provide possible explanations of resulting baryon asymmetry in
cosmology without using the Sakharov's conditions \cite{DiGrezia:2005yx}.

In recent years many aspects of $DSR$ models have been studied
and  some of them are still the subject of
research \cite{Smolin:2010xa}, however there is no consensus about 
which is the position space associated with these theories.

In this work we present a method to construct some DRS models, applying deformed
scale transformations to the  Lorentz generators. In doing this,
a non-linear representation of the Lorentz group arises which, in
turn, leaves the Heisenberg algebra unchanged. The scale
transformations are introduced by deforming a momentum-dependent scale
parameter.  These transformations are carried out on the quantum
operator instead of  on the eigenvalues, as it is usually presented in the
literature \cite{Heuson:2003zt}. We show the equivalence of these two
approximations.  

This work is organized as follows: in Sec. II some
generalities about the Lorentz group are presented. In section III the
deformed scale transformations of the momentum and position operators
are calculated and in section IV the non linear Lorentz transformation
is applied to both the momentum and position operators.  The new dispersion
relations are found and the two scales are shown explicitly for these
theories. Finally, in section V the equivalence between the
transformation on the momentum operators and on the momentum of the
particles is proven and some particular cases are analyzed. 


\section{General Considerations}
It is well known that the angular momentum operators can be written as a function of  
the momentum and position operators as follows  
\be\label{M} M^{\alpha\beta}=i (p^\alpha x^\beta -p^\beta
x^\alpha). \ee 
They are the boosts and rotations  generators  and 
satisfy  the Lorentz algebra
\begin{eqnarray}\label{MM}
  &\left[ M_{\alpha\beta}, M_{\mu\nu}\right] =\nonumber \\
 & -i\left(g_{\alpha\nu} M_{\beta\mu}+g_{\beta\mu}
    M_{\alpha\nu}-g_{\alpha\mu} M_{\beta\nu}- g_{\beta\nu}
    M_{\alpha\mu} \right)\, ,
\end{eqnarray}
this  algebra is a consequence of the more fundamental Heisenberg algebra
\be\label{px}[x^\alpha,x^\beta]=0,\,\,\, [p^\alpha,p^\beta]=0,\,\,\,
[p^\alpha,x^\beta]= ig^{\alpha\beta}.\ee \noindent  
The set of $M$ matrices in (\ref{M}) are the generators of an unitary
representation of the Lorentz group. 
\begin{equation}
\Lambda= \exp (i \omega_{\alpha\beta} M^{\alpha\beta}/2).
\end{equation}

We can find other
unitary  representations for the Lorentz group applying an unitary
transformation to the group elements
\begin{equation}
\Lambda \rightarrow \widetilde\Lambda= U \Lambda U^{\dagger}.
\end{equation}
which implies the  transformation on the generators
\begin{eqnarray}\label{UMU}  
M \rightarrow {\widetilde M}^{\alpha \beta}=U \,M^{\alpha\beta}U^{\dagger}\,. 
\end{eqnarray}
these new generators satisfy the same Lorentz algebra (\ref{MM}).

It is clear that if $U$ and the $M$'s commute, these representations
are  actually the same. Thus,  in order 
to find other non trivial equivalent representations, $U$ must not be a Lorentz scalar.
That transformation on the generators is carried out by the transformation on the basic operators
\begin{eqnarray}\label{Up}
p^\alpha \rightarrow \widetilde p^\alpha = U p^\alpha  U^{\dagger},\\[2mm]\label{Ux}
x^\alpha \rightarrow \widetilde x^\alpha = U x^\alpha  U^{\dagger},
\end{eqnarray}
which, as stated, preserve the Heisenberg algebra.
We can build non standard representation of the Lorentz group using an adequate operator $U$. 
In this paper we will construct such representation, starting with
the ordinary scale transformation on the basic operators but using 
a parameter that depends on some components of the momentum operator 
in order to prevent $U$ from being a Lorentz scalar.
This is not a new idea; in fact the Magueijo-Smolin model \cite{Smolin:2006pa}
is generated in this way, but  constructed   in the momentum representation.
We will extend that procedure for the quantum field operators.

\section{Deformed Scale Transformations}

Let us now  build the ordinary scale transformations in order to
gain the  necessary insight and learn how they can be deformed. 
The finite scale transformation is given in equation (\ref{Up}) with
$U=\exp(\epsilon D)$ are 
 \be\label{UpU} \widetilde p^\alpha=
e^{\epsilon D} p^\alpha e^{-\epsilon D}, \ee \noindent
where $\epsilon$ is a  parameter and $D$ is the dilatation
operator given by \be\label{Dilaton}D= i p_\alpha x^\alpha + c,\ee
with $c$ some constant. The Heisenberg algebra (\ref{px}) implies \be
[D,p^\alpha]=p^\alpha,\,\,\, [D,x^\alpha]=-x^\alpha.\ee
The ordinary  scale
transformations mean that $\epsilon$ is a constant and the transformation
(\ref{UpU}) can be reduced to $ \widetilde
p^\alpha=e^\epsilon p^\alpha;$ in this case and since
$M^{\alpha\beta}$ commutes with $D,$ we have $\widetilde \Lambda= \Lambda$.

In order to avoid the usual scale transformation one can propose that
$\epsilon$ does not commute with the $M^{\alpha\beta}$, although it should indeed
commute with $p^\alpha$.  This can be done by choosing $\epsilon$
as a function of $p$ and not Lorentz invariant. In this
case we obtain one class of deformed scale transformations.

We will work with the scale parameter $\epsilon$ as an homogeneous
function of degree $s$ in $p$, that is
\be\label{eap} \epsilon (ap)=a^s \epsilon ( p),\ee 
where $a$ is a constant. Now $ \epsilon
$ has dimension $s$ and we have \be\label{De} [D,\epsilon]=
s\epsilon,\ee \noindent where the operator $D$ is given in (\ref{Dilaton}) with
Re $ \, c= 2+s/2$, in order to make $\epsilon D$ anti Hermitian or
equivalently, $U$ unitary.

\subsection{Deformed Scale Transformation of $p$ and $x$}

The expression (\ref{UpU}) can be written, using a Hausdorff expansion, as
\be\label{pe} 
\widetilde p^\alpha
=\sum_{n=0}^\infty \frac{[\![(\epsilon D)^{(n)}, p^\alpha]\!]}{n!} ,
\ee \noindent where $ [\![\dots ]\!]$ is the multiple commutator
defined by the recurrence relation
\begin{equation}\label{conmutator} [\![A^{(n+1)}, B]\!]\equiv [ A,
  [\![A^{(n)}, B]\!]],\end{equation}
\noindent with the initial condition $[\![A^{(0)}, B]\!]=B$. 
Proposing $[\![(\epsilon D)^{(n)}, p^\alpha]\!]= \theta_n \epsilon^n p^\alpha$ and using 
(\ref{conmutator}) we find $\theta _{n+1}=(ns+1)\theta_n$, 
with the initial condition $\theta_0=1$, this gives 
\be\label{edp}  \theta_n=  (-1)^n  s^n
\frac{(-1/s)!}{(-1/s-n)!}\, .\ee \noindent 
Introducing (\ref{edp}) in
(\ref{pe}) and adding all the terms, we have
\begin{equation}\label{tildep} \widetilde
  p^\alpha=(1-s\epsilon)^{-1/s} p^\alpha . \end{equation}
\noindent In the simplest case, $s=0$, we obtain $\widetilde p^\alpha=
e^{\epsilon} p^\alpha$ as expected when $\epsilon $ is a
constant; nevertheless in (\ref{tildep}) with $s=0$ $\epsilon$ can be a  function of $p$. 

Similarly, the  transformation of  $x$ is performed via a
Hausdorff expansion for (\ref{Ux}) which now reads
\be\label{xe}
\widetilde x^\alpha
 =\sum_n \frac{[\![(\epsilon D)^{(n)}, x^\alpha]\!]}{n!}. 
\ee 

\noindent
In general the $n$-th
commutator can be parametrized as \be\label{edx} [ \![(\epsilon
D)^{(n)}, x^\alpha ] \!  ]= \alpha_n \epsilon^n  x^\alpha+
i\beta_n\epsilon^{n-1}\epsilon ^\alpha D,\ee 

\noindent 
where 
\be\label{ea}
\epsilon^\alpha=i[x^\alpha,\epsilon]=\frac{\partial
  \epsilon(p)}{\partial p_\alpha}.\ee \noindent 
From
(\ref{conmutator}) in (\ref{edx}) we obtain the recurrence relation
\[ \alpha_{n+1}=(ns-1)\alpha_n,\]
\[ \beta_{n+1}=[(n-1)s-1]\beta_n+\alpha_n,\] to finally find
\be\label{a}\alpha_n= s^n\frac{(1/s)!}{(1/s-n)!}(-1)^n ,\ee \noindent
\be\label{b} \beta_n=n\alpha_{n-1},\ee \noindent where the initial conditions 
$\alpha_0=1$ and $\beta_0=0$ were considered. 

Adding all the terms
in (\ref{xe}) and considering (\ref{a}) and (\ref{b}) we obtain
\begin{equation}\label{tildex} \widetilde
  x^\alpha=(1-s\epsilon)^{1/s}[x^\alpha+i\epsilon^\alpha
  D].\end{equation}

\noindent 
One  can check from equations (\ref{tildep}) and (\ref{tildex}) that the
canonical relations
$$[\widetilde p^\alpha,\widetilde
x^\beta]=ig^{\alpha\beta}$$ still hold.


\subsection{Non Linear Lorentz Transformations.}

Using the expressions for the scaled momenta (\ref{tildep}) and
coordinates (\ref{tildex}), one can construct explicitly the new
Lorentz generators from (\ref{UMU}),

\begin{eqnarray}
  \widetilde M^{\alpha\beta}= M^{\alpha\beta} - i(p^\alpha\epsilon^\beta-
  p^\beta\epsilon^\alpha)D.
\end{eqnarray}
The non-linear Lorentz transformations over the momentum operators are
therefore
\be p^\alpha\rightarrow \widehat p^\alpha= \widetilde\Lambda^{\dagger}
p^\alpha\widetilde\Lambda.\ee
Associativity of this transformation can be implemented in steps, 
$\widehat p^\alpha= U(\Lambda^{\dagger}(U^{\dagger} p^\alpha
U)\Lambda) U^{\dagger},$
to give 
\be\label{hatp} \widehat
p^\alpha =[1+s(\epsilon'-\epsilon)]^{-1/s} p^{\alpha'},\ee \noindent
where $\epsilon'$ is the $\epsilon$ function applied over
$p^{\alpha'}={\Lambda^{\alpha'}}_\beta p^\beta$.  From (\ref{hatp})
it can be seen that if $\epsilon$ is a Lorentz scalar, then
$\epsilon=\epsilon'$ and $\widehat p^\alpha=p^{\alpha'}$.
 
In the same fashion, as in the case of $p$, under non-linear Lorentz
transformation, $x$ transforms as \be\label{hatx} \widehat
x^\alpha=[1+s(\epsilon'-\epsilon)]^{1/s}
[x^{\alpha'}-i({\epsilon'}^{\alpha'}-\epsilon^{\alpha'})D
],\ee \noindent where
$\epsilon^{\beta'}={\Lambda^{\beta'}}_\alpha \partial\epsilon/\partial
p_\alpha$ and $\epsilon'^{\beta'}=\partial\epsilon'/\partial
p_{\beta'}$.

Once again, from (\ref{hatp}) and (\ref{hatx}), it follows that
the new operators satisfy the canonical commutation relations
\[ [\widehat p^\alpha,\widehat x^\alpha]=ig^{\alpha\beta}.\]

\section{General Properties of the Non-Linear Lorentz Transformation.}

A simultaneous unitary transformation over the momentum operators $p$
and coordinates $x$ can be called a canonical transformation, 
just like in classical mechanics. This is  
because it preserves the canonical commutation relations (\ref{px}).  This
type of transformation can be seen as a pasive canonical transformation since
the coordinate operators in the phase space are transformed  
and  the coordinates themselves 
are not. The corresponding active transformations
occur  when only the states are
directly transformed.

Let the state $|p_e\rangle$ be an eigenstate of $p^\alpha$ with
eigenvalue $p_e^\alpha$, 
\be p^\alpha |p_e\rangle =p^\alpha_e
|p_e\rangle.\ee 
under non-linear Lorentz transformations, it
changes as 
\be |p_e\rangle \rightarrow \widetilde
\Lambda|p_e\rangle. \ee 

\noindent To find out how a new state appears we apply $p^\alpha$ as in  
\be p^\alpha \widetilde
\Lambda|p_e\rangle= \widetilde \Lambda( \widetilde \Lambda ^\dagger
p^\alpha \widetilde \Lambda)|p_e\rangle= \frac{p_e^{\alpha'}}{[1+s
  \epsilon(p_e')-s\epsilon(p_e)]^{1/s}}\widetilde
\Lambda|p_e\rangle,\ee 

\noindent which means that
$\widetilde\Lambda|p_e\rangle$ is an eigenstate of $p^\alpha$ with
eigenvalus $\widehat p_e^\alpha =p_e^\alpha /[1+s
\epsilon(p_e')-\epsilon(p_e)]^{1/s}$. Thus  one can write
\be\widetilde\Lambda|p_e\rangle=|\widehat p_e\rangle.\ee 

\noindent The
equivalence of active and passive transformations  is realized as
the  invariance of the mean value of $p$ when the transformation is carried out, 
\be\langle\psi|p\,|\psi\rangle\rightarrow \langle\psi|\widehat
p\,|\psi\rangle= \langle\widehat \psi|p\,|\widehat
\psi\rangle.\ee

According to (\ref{hatx}) we can see that it is not possible to 
proceed similarly  for the coordinates because the operator 
transforms  mixing  the coordinate and  momentum operators.  

An apparent paradox is found here since coordinate transformations can be
written in such a way that the particle spacial coordinates are now a
function of its energy. This can be explained by considering that,
from the operator point of view, the new spacial coordinates depend on
the old momenta, so the new and old coordinates do not actually
commute, although the new momenta do commute with their untransformed
partners.

A point in space time is the eigenvalue of the $x$ eigenstate, which
describes a particle with a well defined position. This state, as
seen by any other observer, is a superposition of the $|\widehat x \rangle$
eigenstates, so it does not have a well defined position in the new
system. 

\subsection{Velocity Scale.}

From now on we will work with the eigenvalues of the momentum
operators instead of the operators themselves and we will omit the subindex $e$.
From (\ref{hatp}) we conclude that  the moment eigenvalues satisfy
\be\label{p/p}\frac{\widehat{\bf
    p}}{\widehat p^0}=\frac{{\bf p}'}{ p^{0'}}.
\ee 

\noindent Calling ${\bf p}/p^0$ the Lorentz velocity $\bf v$, we
have  $ \widehat{\bf v}={\bf v}'$, from the linear Lorentz
transformation (\ref{p/p}) can be written
in terms of the velocities as \be\label{vpvp} \widehat v_{_\parallel}
=\frac{ \beta -v_{_\parallel} }{ 1 - \beta v_{_\parallel}}, \,\,\,
\widehat v_{_\perp} =\frac{ v_{_\perp} }{ \Gamma ( 1 - \beta
  v_{_\parallel})}, \ee \noindent where $\perp $ and $\parallel $
stand for the parallel and perpendicular components of the Lorentz velocities
with respect to the relative velocity $\beta$, and
\be \Gamma=\frac{1}{\sqrt{1-\beta^2}}.\ee 

\noindent  
This transformation satisfies the addition rule as in the conventional
relativity, thus the speed of light $v=1$ is still a natural scale of
the theory.  To see that more clearly one can write (\ref{vpvp}) as
\be\label{vv} 1-\widehat v^2= \frac{1}{ \Gamma^2(1-\beta v_{\parallel
  })^2} (1-{v}^2).\ee 

\noindent From (\ref{vv}) we see that if
$\beta<1$ and $v\leq 1$ then $\widehat v\leq1$. But the Lorentz velocity is not the 
real particle velocity, it is only the velocity of the particle in the 
limit $\epsilon \rightarrow 0$. The connection between Lorentz velocity and particle velocity will be 
developed in the next subsection.

\subsection{Dispersion Relations.}

The function $\epsilon(\widehat p)$ is given by 
\be\label{hate}
\widehat\epsilon\equiv\epsilon(\widehat
p)=\frac{\epsilon'}{1+s(\epsilon'-\epsilon)},\ee

\noindent where the
homogeneity of $\epsilon$   (\ref{eap}) was used. From this expression and
(\ref{hatp}) one can obtain
\be\label{p/e} \frac{\widehat p^\alpha
}{{\widehat\epsilon}^{\,1/s}}=\frac{
  p^{\alpha'}}{{\epsilon'}^{1/s}}.\ee

\noindent
Moreover, from (\ref{hate}) we get
\be
\frac {\widehat \epsilon}{\epsilon'} =\frac {1-s\widehat
  \epsilon}{1-s\epsilon},\ee 

\noindent therefore (\ref{p/e}) is written as
\be\label{p/1-e} \frac{\widehat p^\alpha}{(1-s\widehat
  \epsilon\,)^{1/s}} = \frac{ p^{\alpha'}}{(1-s \epsilon)^{1/s}}.
\ee 

\noindent
Squaring both sides of (\ref{p/1-e}) we find an invariant quantity,
which can be identified as the invariant mass of the particle
\be\label{m} \frac{ p^2}{(1-s \epsilon)^{2/s}}=m^2. \ee 

\noindent This
equation also gives us the new dispersion relations; that is, a new relation between  momentum and  energy.

Despite some discussions in the literature on a proper definition of particle speed in the DSR theories,
\cite{Kosinski:2002gu}, \cite{Ghosh:2007rw}, in this paper we 
take  the  particle velocity as the group velocity $ {\bf u}={\partial p_0}/{\partial{\bf p }}$ 
which is different from  the Lorentz velocity ${\bf v}$. 
Starting from (\ref{m})  and taking the $ {\bf p }$ derivative  we find
\be \label{vp}
{\bf u}={\bf v}  \left(\frac{1-s\epsilon + p^i\epsilon _i /(v\gamma)^2 }{1- s\epsilon -p^0 \epsilon _0/\gamma^2 }\right),
\ee
where $i=1,2,3$ and $\epsilon _\mu= \partial \epsilon/\partial  p^\mu$. 
One can see  that if $|\bf v|\rightarrow 1$ then  
$1/\gamma^2 \rightarrow0$ and $|{\bf u}|\rightarrow 1$. 
This means that the particle velocity has the same limit that the Lorentz velocity.

At this point we can calculate all the dynamics of the free particle 
starting from the  Hamiltonian, which is   given by $H= \int {\bf u} \cdot d{\bf p} =p^0$;  
nevertheless, in what follows we will use   
the Lagrangian formalism in covariant form, which in principle is equivalent.

\subsection{Energy Scale}
We will now show that these type of models have both a momentum and an energy scale. 
Let us start analyzing first the massless particles, 
that is $|{\bf v}|=1$ or $|{\bf p}|=p_0$. With this in mind and reversing (\ref{p/1-e}) we  can compute 
\be\label{1/p} \frac{1}{(\widehat p^0 )^s}- \frac{s\widehat \epsilon}{(\widehat
  p^0 )^s} = \frac{1}{\Gamma^s(1-\beta v_{_\parallel })^s}\left( \frac{1}{
    (p^0) ^s}- \frac{s \epsilon}{ (p^0) ^s}\right) 
\ee

\noindent
and because $\epsilon'= \epsilon(\Lambda p)$ and a Lorentz transformation with $p^0=| {\bf p}|$ we have
\[ \epsilon'=  \epsilon\Big(\Gamma (1-\beta v_{_\parallel })\Big)= \Gamma^s (1-\beta v_{_\parallel })^s\epsilon,\] 
we obtain 
\be \label{e/p}
\frac{{\epsilon'}^{1/s}}{p^{0'}}=\frac{\epsilon^{1/s}}{p^0}.\ee

\noindent
According to (\ref{p/e}) and (\ref{e/p}) one can see that
$\frac{\epsilon^{1/s}}{p^0}$ is an invariant for a massless particle; 
this  quantity  has length units and therefore
can be equated to some length $l_{p}$ that could be the same order of the Planck
length
\be\label{lp}l_{p} =\frac{(s\epsilon)^{1/s}}{ p^0 },\ee 
thus (\ref{1/p}) is written as  
\be \frac{1}{(\widehat
  p^0 )^s}-l_p ^s = \frac{1}{\Gamma^s(1-\beta v_{_\parallel }
  )^s}\left( \frac{1}{ (p^0) ^s}- l_p ^s \right). \ee 

\noindent 
If the particle energy in some system satisfies $p_0 < 1/l_p$ then, 
in any other system, the energy will satisfy $\widehat p^0 < 1/l_p$,  
because $\Gamma(1-\beta v_{_\parallel }) >0$.

For massive particles the energy scale is the same. 
It suffices to note that when 
the particle energy is increasing, the Lorentz velocity limit is 1. 
 In this limit the massive particle behaves like a massless particle.
 Then the analysis for the massless particle applies in this limit as well.
 From (\ref{lp}) we can see that for high energy massive 
particles  the quantity $s\epsilon$ behaves like $(l_p p_0)^s$. Hence it is natural to think that, 
for low energy, $s\epsilon $ has the form $(l_p p_0)^sf(v) $ 
where  $f(v)\rightarrow 1$ when $v\rightarrow 1$.  

\section{Covariant Lagrangian Formulations.}

In this section we will find the Lagrangian for free particles in some of these models. 
From the Lagrangian it could be possible to make and educated guess about 
the way in which the coordinates transform  
in those models.  Actually,  the dispersion relations allow us to find the particle 
energy as a  function of the particle momentum, the only obstacle is that 
we must deal with a not easily solvable algebraic equation. 
Nevertheless when $s$ is small it becomes easier. 
In particular,  we will concentrate in  low   values  $s = 1$ and $s=2$. 
For each value of $s$ we  still can choose the function of the Lorentz 
speed magnitude $v$. 
For simplicity we only consider functions of the type $v^r$ where $0<r<s$ 
then for fixed values of $s$ and $r$ we have
\be s\epsilon = l_p^s p_0^{s-r} |{\bf p}|^r,\ee 
and the dispersion relation is
\be\label{pp2} p^2=m^2(1-l_p^s p_0^{s-r} |{\bf p}|^r)^{2/s}.\ee
Therefore for $s=1, 2$  we have five different models: 
(1,0), (1,1), (2,0), (2,1), (2,1) according to $(s,r)$ as in Table \ref{lagrangian}.
Here we can identify the model (1,0) with the Magueijo-Smolin Model \cite{Smolin:2006pa}

For each of these models we need to  find the conjugate 
momentum coordinates and the Lagrangian based on the dispersion relations associated 
with  each model, following the procedure given in \cite{Ghosh:2007rw}. 
The idea is to construct the Lagrangian linearly in the velocities, 
imposing the dispersion relation as a constraint through a Lagrange multiplier in the following way
\be L= \dot x\cdot p -\frac{e}{2}[p^2-m^2(1-s\epsilon)]^{2/s}\ee
then, applying the Euler-Lagrange equations for $p$ and using the 
constraint we finally  find the Lagrangian as a function of the velocities. 
The results of this procedure  for some  models  are shown in the Table \ref{lagrangian}.

\begin{table}[th]
\caption{Lagrangian of the different models.}
\label{lagrangian}
\begin{center}
\begin{tabular}{c|l}
\hline\hline
$(s,r)$& Lagrangian\\
\hline
(1,0)& $ \displaystyle {L=\frac{m}{1-m^2 l_p^2}\left( \sqrt{ \dot x_0^2 -(1-m^2l_p^2) \dot{\bf  x} ^2} - m l_p \dot x_0\right) }$\\[4mm]
\hline
 (1,1)&  $ \displaystyle  L=\frac{m}{1+m^2 l_p^2}\left( \sqrt{ (1+m^2l_p^2)\dot x_0^2 - \dot{\bf  x} ^2} - m l_p |\dot {\bf x}|\right) $\\[4mm]
\hline
(2,0)&   $ \displaystyle  L=\frac{m}{\sqrt{1+m^2 l_p^2}} \sqrt{ \dot x_0^2 - (1+m^2l_p^2)\dot{\bf  x} ^2}  $\\[4mm]
\hline
(2,1)&   $ \displaystyle  L=\frac{m}{{1+ m^4 l_p^4/4}} \sqrt{ \dot x_0^2 -m^2 l_p^2 |\dot{\bf  x}| \dot x_0 - \dot{\bf  x} ^2}  $\\[4mm]
\hline 
(2,2)&  $ \displaystyle  L=\frac{m}{\sqrt{1+m^2 l_p^2}} \sqrt{ (1+m^2l_p^2)\dot x_0^2 - \dot{\bf  x} ^2} $ \\[4mm]
\hline
\end{tabular}
\end{center}
\end{table}


\subsection{Efective Metric.} 
Except in the case (1,1) these  results suggest that the Lagrangians of 
those models  can be written as a function of the effective metric $\tilde g_{\mu\nu}$, to give
\begin{equation} L=m'\sqrt{\dot x^\mu \widetilde g_{\mu\nu}\dot x^\nu}\end{equation}
where $m'$  is chosen in such a way that $\widetilde g_{ij}=-\delta_{ij}$. 

In general $\tilde g_{\mu\nu}$ will depend on the mass of the particle and could  depend 
on the direction of the vector over which  it acts.
We can propose  a Lorentz transformation that leaves invariant this metric,  
$\widetilde\Lambda^T\widetilde g\widetilde\Lambda= \widetilde g$,
writing $\widetilde g=\Gamma^T g \Gamma$  we find that $\widetilde \Lambda =\Gamma^{-1}\Lambda \Gamma$. 

For Example in the models (1,0), (2,0) and (2,2)  
the metric and the $\Gamma$ matrix can be written as  
\begin{equation} \widetilde g=\left(\begin{array}{cc}1/ \alpha^2&0\\0&-1 \end{array}\right) ,\,\, \Gamma=\left(\begin{array}{cc}1/ \alpha&0\\0&-1 \end{array}\right) \end{equation}
where  $m'$ and $\alpha $ are given in the Table \ref{constants}.

\begin{table}[th]
\caption{Constants of the different models.}
\label{constants}
\begin{center}
\begin{tabular}{c|c|c}
\hline\hline
$(s,r)$&$m'$& $\alpha$ \\
\hline
(1,0)& ${m}{(1-m^2l_p^2)}^{1/2}$ & $ (1-m^2l_p^2)^{1/2}= (1+m'^2l_p^2)^{-1/2} $\\[4mm]
\hline
(2,0)&  $ m$  & $(1+m^2l_p^2)^{1/2} =(1+m'^2l_p^2)^{1/2}$ \\[4mm]
\hline
(2,2)&  $ {m}{(1+m^2l_p^2)}^{-1/2}$  & $  (1+m^2l_p^2)^{-1/2} =(1-m'^2l_p^2)^{1/2}$ \\[4mm]
\hline
\end{tabular}
\end{center}
\end{table}

The representation in the  coordinate space  of a Lorentz transformation is
\begin{equation}\widetilde \Lambda= \left(\begin{array}{cc} \gamma& -\gamma 
\alpha{\bf \beta}\cdot\\-\gamma {\bf \beta}/\alpha&\gamma {\rm P_\parallel }+{\rm P_\perp } 
\end{array}\right) ,\end{equation}
where ${\rm P_\parallel }=\beta\beta^T$ is parallel to the velocity 
projection  operator and ${\rm P_\perp }=1-{\rm P_\parallel }$ is the corresponding perpendicular
 projection  operator.
This leads to the following velocity addition formula (in two dimension for simplicity)
\be \label{uu}
u'= \frac {u+ \beta / \alpha}{1+u \beta\alpha},
\ee
this implies that $u'$  depends  on the mass of the particle.  
Note that for  a photon  in any system $u=1$,   this is because   the photon is massless and $\alpha=1$. 
In the other hand (\ref{uu})  implies that ``together'' is a relative concept because the coordinates of particles of different mass transform  differently.

Finally  the particle  momentum $p_\mu=\partial L/\partial {\dot x}^\mu$  satisfy the constraint 
\be\label{pgp} p_\mu \widetilde{g}^{\mu\nu} p_\nu= \alpha^2 p_0^2 -{\bf p}^2= m'^2.\ee
It is easy to prove  that this relation is equivalent to the dispersion relations given in  
\ref{pp2} for the three models with the respective constants shown in Table \ref{constants}.

\subsection{Quantum Field Theory.}

Something  interesting in this kind of theories is the  
fact that there  exists a natural  Lorentz invariant cut-off in the loop integrals  
which appear to higher order in the quantum corrections. 
Let us first consider a  real field scalar associated with a  free particle. In order to quantize the theory the replacement 
$p_0  \rightarrow i\partial /\partial t$ and ${\bf p}\rightarrow-i\nabla$, ($m'\rightarrow m$) are made  in (\ref{pgp}) and 
the modified Klein-Gordon equation reads 
 \begin{equation} \left( {\alpha^2}{\partial _0^2}- \nabla ^2+ m^2 \right)\phi =0.\end{equation}
The Lagrangian for free particle associated to this equation would be 

\begin{equation}  L= \frac{1}{2} \left( \alpha^2 \dot\phi^2  - (\nabla\phi)^2 +m^2\phi^2 \right), \end{equation}
then momentum of the  associated field  $  \pi(x)  = \alpha^2 \dot\phi(x)$ which satisfy the canonical equal time
commutations relations  $[\phi (x) , \pi(y)]_{x_0=y_0}={i} \delta^3({\bf x-y})$.

The Fourier expansion of the Klein-Gordon field is therefore
\begin{equation}  \phi(x)=\int \frac{d^3 {\bf k}}{(2\pi)^3 2\alpha \omega_{\bf k}} \left( a({\bf k}) e^{-i\check{k}\cdot x } +a^\dagger ({\bf k}) e^{i\check{k}\cdot x }   \right) , \end{equation}
with  $\check k= (\omega_{\bf k}/\alpha, {\bf k})$ and $\omega_{\bf k} = \sqrt{{\bf k}^2+ m^2 }$, 
where  the $a$  operator  algebra is given by 
\begin{equation}  [ a({\bf k}),a({\bf k'})] = 2\alpha \omega _ { \bf k }  \delta^3({\bf k-k'}).\end{equation}
Finally  the propagator, in the momentum space has the form 
\begin{equation}  \Delta (k) =\frac{i}{\alpha^2 k_0^2- {\bf k}^2- m^2 },\end{equation}
where Im $ m^2\rightarrow 0^-$.

When the  interaction $\lambda \phi^4$ is introduced the   
 one loop  correction to the scalar propagator is 
\begin{equation} -i\Sigma(k)= \frac{\lambda}{ 8\pi^3} \int_0^{E_p} {\bf k}^2d|{\bf k}|\int_{-E_p}^{E_p} dk_0 \frac{1}{\alpha^2 k_0^2- {\bf k}^2- m^2 }\end{equation}
When this  integral  is carried out 
the leading terms will contain terms proportional to $\sim E_p^2$ and $\sim \ln (E_p/m)$. 
Here $E_p$ acts  a cutoff. This is somehow  natural  because the theory is 
invariant under this deformed Lorentz transformations. Nevertheless we would 
need a finite renormalization term  in order to remove the quadratic and logarithmic terms in $E_p$. 
At this point
 we can not  proceed any further with the radiative corrections. This is because it is not known how 
to add two momenta in order to find the vertex  correction in this new relativity.
This kind of problem is highly non trivial and is  characteristic of all of these theories with two scales.   

\section{Conclusions.}
An alternative approach to the non-linear representations of Lorentz transformations has been introduced; in this approach one changes the generators of a scale deformed transformation as in equation (\ref{UMU}). The deformation of the transformation is obtained using momentum operator-dependent parameters. When the transformation of the operator is calculated in terms of momentum eigenstates it becomes clear that this type of representations of the Lorentz Group correspond to a double scaled invariant models. 
We have found Smolin's model is one of a family of such models, described here. Moreover, this type of transformations can be applied to the position operator and it was found that it transforms mixing both momentum and position operators, very much like a typical canonical transformation in classical mechanics. From this one can conclude that it is not possible to find a coordinate transformation for a particle in this model which does not contain the momentum operator. The interpretation is that having a precise position in space depends on the frame of the observer. If a particle appears at a definite position for one observer (occupying an eigenstate of the position operator in that frame) it appears for another observer, in relative movement, as having a position which is the combination of different eigenstates of the position operator.
A covariant Lagrangian formulation of this relativistic models has been presented as well.  This allows us to study this models introducing an effective metric which has a dependency on the mass of the particle. Advancing in this program we present a second quantization of a spinless particle in this conditions, invariant under the transformations found above. This theory, in principle, should be finite since the integrals in the momenta can be properly re-normalized. This is possible because the second scale of the theory implies a cut-off for the momenta of the particles in the quantized version of the model. Although there appear some quadratic terms on the energy scale they could be removed as well using renormalization again. Vertex corrections are a problem since the addition rules for the momenta in these models is not well understood and therefore vertex corrections are yet to be calculated.  
Double scaled Lorents transformations are an interesting approach to understand a fundamental length scale in nature. This work advances that understanding and proposes a whole family which systematically introduces those models, although several problems persist.  More research would be needed to find for example non-linear Lorentzian transformations extended to include more parameters in the scale transformations that can depend on the coordinates. With this one would obtain a length, instead of an energy, cut-off.

\section*{Acknowledgments.}

This paper was supported by the \textit{Comit\'e para el Desarrollo de la Investigaci\'on} CODI, Universidad de Antioquia.

\end{document}